\documentclass[12pt,preprint]{aastex}
\newcommand{\gcm}{g~cm$^{-3}$}
\shorttitle{X- and Gamma-Ray Flash Signatures from SNeIa?}
\shortauthors{H\"oflich \& Schaefer}
\begin{document}
\title{X- and Gamma-Ray Flashes from Type Ia Supernovae?}
\author{
Peter H\"oflich\altaffilmark{1} and
Bradley E. Schaefer\altaffilmark{2}
}
\altaffiltext{1}{Dept. of Physics, Florida State University, 315 Keen Building, Tallahassee, Florida, 32306-4350}
\altaffiltext{2}{Department of Physics and Astronomy, Louisiana State University, Baton Rouge, Louisiana 70803}

\begin{abstract}

We investigate two potential mechanisms that will produce X-ray and $\gamma$-ray flashes from Type Ia supernovae (SN-Ia).
  The first mechanism is the breakout of the thermonuclear burning front as it reaches the surface of the white dwarf.
    The second mechanism is the interaction of the rapidly expanding envelope with material within an accretion disk 
    in the progenitor system.  Our study is based on the delayed detonation scenario because this can account for the
     majority of light curves, spectra, and statistical properites of 'Branch-normal' SN-Ia.  Based on detailed 
     radiation-hydro calculation which include nuclear networks, we find that both mechanisms produce brief 
     flashes of high energy radiation with peak luminosities of $10^{48}-10^{50}$ erg/sec.  The breakout
     from the white dwarf surface produces flashes  with a rapid exponential decay by 3 to 4 orders of magnitude 
     on time scales of a of a few tenths of a second and
     with most of the radiation in the X-ray and soft-$\gamma$-ray range.
            The shocks produced in gases in and
      around the binary will produce flashes with a characteristic duration of a few seconds with most of the 
      radiation coming out as X and $\gamma $-rays.  In both mechanisms, we expect a fast rise and slow decline and,
      after the peak, an evolution from hard to softer radiation due to adiabatic expansion.  In many cases, 
      flashes from both mechanisms will be superposed.  The X- and $\gamma$-ray visibility of a SN-Ia will 
      depend strongly on self absorption within the progenitor system, specifically on the properties of the 
      accretion disk and its orientation towards the observer.  Such X-ray and $\gamma$-ray flashes could be detected as 
      triggered events by Gamma-Ray Burst (GRB) detectors on satellites, with events in current GRB catalogs.  We 
      have searched through the GRB catalogs (for the BATSE, HETE, and Swift experiments) for GRBs 
      that occur at the extrapolated time of explosion and in the correct direction for known Type 
      Ia supernovae with radial velocity of less than 3,000 km s$^{-1}$.  
       For BATSE about $12.9\pm 3.6$ nearby SNe~Ia should have been detected, but only $0.8 \pm 0.7$
       non-conincidental matches have been found.
       With the HETE and Swift satellites, we expect to see $5.6 \pm 1.3 $SN-Ia flashes from known nearby SNe~Ia
       but, yet, no SN-Ia flashes were detected.
       With the trigger thresholds for these experiments and the upper limits on the SN-Ia distances, 
       we show that the bolometric peak luminosity of SN-Ia Flashes must be less
        $\sim 10^{46}$ erg s$^{-1}$.  Our observational limit is several 
        orders-of-magnitude smaller than the peak luminosities predicted for both the early flash. We
        attribute this difference to the absorption of the X- and $\gamma$-rays by
        the accretion disk of large scale height or common envelope that would be smothering the white dwarf.

\end{abstract}
 

\section{Introduction}

	The history of Gamma-Ray Bursts (GRBs) started out with a supernova (SN, SNe plural) connection 
when Stirling Colgate calculated that the shock breakout of a Type II SN should create a burst of 
gamma radiation (Colgate 1968; 1970; 1974), and then he asked the Los Alamos Vela group to see if 
they could recognize such events.  Indeed, Klebesadel, Strong, \& Olson (1973) discovered the 
Gamma-Ray Burst phenomenon, although it was quickly realized that the shock breakout from Type
 II SN would not occur at gamma ray photon energies.  
	
	GRBs must necessarily pack large amounts of energy in a small volume, so attempts to link GRBs 
and SNe have persisted since 1973.  From 1979 until the 1990's, a strong link (Felton 1982) was 
provided by the unique and bright burst seen on 5 March 1979 (Cline et al. 1980) coming from near 
the middle of a supernova remnant in the Large Magellanic Cloud (Evans et al. 1980).  However, we 
now realize that this event is a separate subclass of bursts, called the Soft Gamma Repeaters, 
that apparently are magnetars and completely separate from the classical GRBs (Hurley 2000).
	
	The first strong SN/GRB connection was made when the burst GRB 980425 was found to have a 
coincidence in time and position to SN 1998bw (Galama et al. 1998).  The GRB was lower in 
luminosity than other known GRBs by many orders of magnitude and the SN was a highly unusual 
Type Ic SN with very high expansion velocities and a record breaking radio luminosity 
(Kulkarni et al. 1998).  So both GRB and SN were so unusual that it was risky to 
generalize the connection to all events.  Over the next few years, various late-time 
bumps in the light curves of burst afterglows have been claimed to be an underlying 
supernova (Bloom et al. 1999; Reichart et al. 1999; Galama et al. 2000), but these 
claims all had poor data and bumps are seen in afterglow light curves on all time 
scales so there is no reason to connect any particular bumps with supernovae.  A 
stronger SN/GRB connection was made with the discovery of high-velocity high-excitation 
absorption lines in the spectrum of GRB 021004 points to the GRB progenitor being a 
Wolf-Rayet star (Schaefer et al. 2003).  Various groups also sought statistical 
connections between SNe and GRBs.  The first claim (Wang \& Wheeler 1998) was that 
bright and well-observed Type Ib/c SNe are statistically correlated with GRBs.  This 
connection has been strongly rejected on statistical grounds (Deng 2001; Schaefer \& 
Deng 2000) as well as through the use of better GRB error boxes (Kippen et al. 1998).  
Soon, claims had been made connecting specific Type IIn SNe with GRBs (Terlevich \& 
Fabian 1999), but these also have low significance (Deng 2001; Schaefer \& Deng 2000). 
 In the meantime, strong theoretical models were being developed which connect 
 long-duration GRBs with the core collapse of very massive stars with fast rotation 
 (MacFadyen, Woosley, \& Heger 2001; Woosley \& Bloom 2006).  In 2003, the HETE2 
 satellite discovered a relatively nearby normal burst (GRB 030329) which displayed 
 an afterglow spectrum like SN 1998bw starting in the week after the burst 
 (Stanek et al. 2003; Hjorth et al. 2003).  Further high-confidence associations 
 between normal GRBs and Type Ic SNe have been made for GRB 031203 and SN 2003lw 
 (Malesani et al. 2004) and for GRB 060218 and SN 2006aj (Campana et al. 2006).  
 With all these strong and weak connections, the community is now confident that 
 almost all the long-duration GRBs are associated with a core collapse supernova 
 explosion.

	 This still leaves open the question of whether the short duration GRBs (Cline 
\& Desai 1974; Kouveliotou et al. 1993) are associated with SNe?  The currently 
popular model is that the short GRBs are caused by the collision of two neutron 
stars in a binary orbit which in-spirals due to gravitational wave emission 
(Taylor 1994).  Many reasonable alternatives have been proposed (Dado \& Dar 2005), 
including carbon-oxygen white dwarf and neutron star mergers (Dar \& DeRujula 2004), 
gravitational collapse of a neutron star to form a quark star (Dar 1999), super-flares 
from Soft Gamma Repeaters in nearby galaxies (Hurley et al. 2005), or just simply some 
variation on the long-duration GRB core collapse.  A substantial advance was made with 
the identification of five x-ray and three optical afterglows associated with short 
duration bursts (Gehrels et al. 2005; Villasenor et al. 2005; Burrows et al. 2005; 
Fox et al. 2005; Hjorth et al. 2005; Soderberg et al. 2006).  These afterglows 
are associated with moderately bright and nearby galaxies, but this must be some sort 
of selection effect as many other short bursts certainly have no galaxy association to 
deep limits (Schaefer 2006).  The three afterglows with optical positions are associated 
with the outer parts of the galaxies and with elliptical galaxies; both of which strongly
 point to the progenitors being in an older population.  In addition, very strong limits 
 have been placed to show that there are no supernovae associated with the bursts (Fox et al. 
 2005; Hjorth et al. 2005; Berger et al. 2005; Bloom et al. 2006).  In all, it does appear 
 that the short duration GRBs are an old population, often do not have an associated 
 supernova, and are a separate population from the long duration GRBs.
	 
	 The purpose of this paper is to examine another connection between GRBs and SNe.  
In particular, we calculate that Type Ia SNe should produce short duration flashes 
of X-rays and $\gamma$-rays that would appear as short duration GRBs and would be 
discovered with past and current GRB detectors.  It is possible that relatively 
nearby SN-Ia events will produce an X-ray or $\gamma$-ray flash that is bright 
enough to be detected. 
%PAH
 It may be worth mentioning that  x/gamma-ray flashes from a GRB-connected
SN (i.e. SN2006aj) might have already been detected in GRB060218,
as suggested by Campana et al. (2006) and Waxman et al. (2007).
 Such flashes might either be labeled as short duration 
GRBs or as X-Ray Flashes (Heise et al. 2001; Kippen et al. 2001).
  We do not 
think that the flashes from SN-Ia events can account for the diversity of 
either the short duration GRBs or the X-Ray Flashes, so we are expecting that 
the SN-Ia flashes are only a subset of the triggered events.
	 
	 Our original motivation for this study was the realization that the 
inevitable shock breakout of a Type Ia event will likely produce a burst 
of X- and $\gamma$-radiation lasting for perhaps seconds of time.
%PAH
  The mechanism is similar to that of the original Colgate proposal for Type 
II SNe, however, with the distinguishing feature
that, initially, the front starts as a weak detonation which is propelled by 
nuclear burning with time scales of seconds, and on an already rapidly expanding
background. The result of the nuclear burning front
should be a heating of the outermost material to a temperature of tens 
of keV that will last for a few seconds until adiabatic cooling (from 
the expansion of the material and balanced by nuclear burning) reduces the temperature.
  During this brief
time interval, the emission will be of hard radiation from a surface
area with a characteristic radius of $ \approx 10^{10}$ cm.  Such a
source would produce a short burst of X and gamma radiation which should be
visible over gigaparsec distances.  To retain a distinction with the 
GRB phenomenon, we will label these events as 'SN-Ia Flashes'.
	 
\section{Radiation Hydrodynamical Models for Thermonuclear Supernovae}
 
The basic explosion mechanism for Type Ia Supernovae is that carbon 
burning in the center of a white dwarf (WD)  leads to a thermonuclear 
runaway because the degenerate electron gas shows hardly any temperature 
dependence, and the energy release results in the explosion.  To first 
order, the outcome hardly depends on details or even the general scenario 
because nuclear physics determines the structure of the WD and the energy 
release, which causes 'stellar amnesia' (H\"oflich et. al 2003). The apparent
homogeneity of SN-Ia events does not imply an unique explosion scenario 
but masks the complexity of a phenomenon which
includes stellar evolution, rotation and mass loss, accretion physics, 
physics of the ignition process, propagation of nuclear flames and 
transport phenomena.

Within this general picture, two classes of models are most likely 
realized: (1) An explosion of a  carbon/oxygen WD with a mass close 
to the Chandrasekhar limit ($M_{Ch}$), which accretes matter through 
Roche-lobe overflow from an evolved companion star (Whelan \& Iben, 1973).
In this case, the explosion is triggered by compressional heating near 
the WD center.  Alternatively, (2) the SN could be an explosion of a 
rotating configuration formed from the merging of two low-mass WDs, 
after the loss of angular momentum due to gravitational radiation 
allows for collapse (Webbink 1994, Paczy\'nski 1985, Benz et al. 1990).

The last decade has witnessed an explosive growth of high-quality 
data which allow study of second order effects.  In combination 
with advances in computational methods, this provided new insights 
into the physics and a link to observations.  The majority of SN-Ia 
seems to originate from the explosion of a WD close to $M_{Ch}$ 
(H\"oflich \& Khokhlov 1996). Based on detailed analyses of light curves and spectra,
the most likely scenario involves an early phase of deflagration 
burning which is followed by a phase of detonation (DDT, see below), 
called delayed detonation models (Khokhlov 1991). An initial
deflagration phase is needed for $M_{Ch}$ mass WDs to allow for 
the production of intermediate mass elements, and a subsequent 
detonation phase is required to be in agreement with the overall 
radially layered chemical structure and the observation that almost 
the entire WD is burned. For recent reviews, see Branch (1999),
H\"oflich (2006),  and  Nomoto (2003).

Here, we want to mention two results directly relevant for  bursts, 
and which set the tone.  In a recent study of early time spectra of 
several SN-Ia, Quimby et al. (2005) established that, as suspected 
(e.g., Branch 1999, H\"oflich 1995, Marion et al. 2003), the nuclear burning front
reaches the very outer layers of SN-Ia where the outer layers of 
SN expand with velocities in excess of 25,000 km/sec.  Secondly, 
high velocity Ca II has been found to be a common feature in SN-Ia 
(Fisher et al. 1999, Wang et al. 2003).  Gerardy et al. (2004) studied the
formation of the high velocity Ca II feature and its diagnostics 
based on detailed NLTE-models.  They showed that this feature and 
its evolution with time can be understood in the framework of the 
interaction of the ejecta with a circumstellar shells of solar 
composition  which, likely, has been part of the progenitor 
system/accretion disk with a dimension of $ 2-10 \times 10^{10}$ cm
(Iben \& Tutukov 1975).  Interaction with a wind was excluded because ongoing
 interaction would dominate the luminosity of SNe~Ia. Quimby et al. (2005)
 applied this diagnostics to several supernovae and estimated the 
 mass of the shell to be between $10^{-3}$ to  $2 \times 10^{-2}M_\odot $.  
  These estimates are consistent with the upper limits based on hydrogen 
  emission by Cunning et al. (1989) and, more recently, Mattila et al.(2005).

Our study is based on the delayed-detonation scenario because it reproduces 
the optical and infrared light-curves and spectra and the statistical 
properties of typical SN-Ia  reasonably well.  During the early phase, 
the flame propagates as a deflagration, i.e. the unburned matter is 
ignited by heat conduction over a  front propagating with  an effective 
velocity of a few percent of the speed of sound.  After burning  
$\approx 0.3 M_\odot $ of the carbon/oxygen WD, the detonation is 
triggered.  In a detonation, the matter is ignited by compression 
and the front is driven by nuclear burning behind the front. 
 
We consider two possible origins for X-ray and $\gamma$-ray flashes.
 The first possibility is the breakout of the (nuclear) burning front
  on the surface of the white dwarf, and the second possibility is the 
  interaction of the rapidly expanding envelope with material in an 
  accretion disk within the progenitor system.  In either case, the 
  high energy ultimately comes from the thermonuclear energy.  The 
  total amount of energy available is determined by the thermal and 
  total energy content of the outer layers in case of the outbreak 
  and interaction, respectively.

\subsection{Numerical Methods and Setup}
 
The computations have been performed using our HYDrodynamical RAdation code (HYDRA) 
which is based on modules used to carry out many  prior studies of SN-Ia.  Previous 
applications include detailed, hydrodynamical calculations including detailed nuclear
 networks, $\gamma $-ray transport in spherical and 3-D geometry, and detailed NLTE 
 light curves and spectra ({H\"oflich  1988, H\"oflich, M\"uller \& Khokhlov 1993, 
 H\"oflich 1995, Howell et al.  2001).  For technical details of HYDRA, see H\"oflich 
 et al. (1998),  H\"oflich (2002ab), and references therein.

Parameters were  chosen, which roughly match the observed properties of normal Type~Ia 
supernovae. 
We consider the explosion of a Chandrasekhar mass white dwarf which originates from a 
star with a main sequence mass of 5 $M_\odot$ and solar composition.  At the time of 
the explosion, the central density is $2 \times 10^9$ \gcm. The nuclear burning starts 
as a deflagration with a parameterized description of rate of burning based on 3-D 
models by Khokhlov  (2001). When the density reaches $\rho_{tr}=2.5 \times 10^7$ \gcm,
the 
detonation is triggered.  Although we consider a specific model, the results are 
more generally applicable since the structure of the WD, the explosion energy, and 
the light-curves are mainly determined by nuclear physics rather than the details 
of the nuclear burning (``stellar amnesia''; H\"oflich et al. 2003).  From the setup,
the model is identical to 5p0z22.25 of H\"oflich et al. (2002) but with some technical 
modifications to allow for this study.  The explosion is calculated for the first 
30 seconds including simultaneously the  hydrodynamics, the nuclear reaction 
networks with 218 isotopes, radiation transport modules that take into 
account relativistic corrections (Mihalas, Kunacz \& Hummer 1976), and the time 
dependence for the radiation transport. We consider 800 frequency groups ranging 
from $10^{-5}$ to 3 $MeV$.  For the opacities, we include bound-free, free-free 
and Compton scattering, pair production and inverse reactions, and nuclear reactions  
(H\"oflich 1991, Hoeflich 2005).  We assume full ionization rather than detailed
atomic models because temperatures are of the order of $10^7~-~10^9$ K.  
Our hydrodynamics code uses a fixed grid with a 
mass resolution of $2 \times 10^{-6} M_\odot $ in the outer layers of the WD.  
For the the radiation transport, we allow for 7 levels of rezoning to increase 
the effective solution by up to a factor of $2^7$, i.e to $2 \times 10^{-8} M_\odot $. 
 Hydrodynamical quantities are interpolated using 'rotated parabolae'.

\subsubsection{The Evolution and the Shock Breakout}

In Fig. 1, we show the structure of the exploding WD during several phases.
  During the subsonic deflagration phase (a), the nuclear energy release causes a 
  pre-expansion of the entire WD from about 1700 km to $\approx$5500  km.  Burning
   in the inner layers pushes the outer layers. Most of the nuclear energy is used
    to lift the WD in its gravitational potential and even at the outer layers, the 
    expansion velocity remain less than a few thousand km/sec even at the outermost 
    layers. During this phase, the temperature of the burned matter reaches well in execess 
    of $5 \times 10^9$ K behind the front, and adiabatic cooling causes the temperature 
    to drop in the unburned region.  Subsequently, a weak detonation front (b) travels 
    through the expanding WD at velocity slightly larger than sound speed and heats the 
    WD to temperatures of a few times $10^{9}$ K.  Nuclear burning behind the front drives 
    the compressional wave and causes the expansion of the matter behind the front at an 
    accelerated rate.  At about 2.3 seconds, the detonation front reaches the surface and 
    heats the outermost layers to $\approx 1.5 \times 10^{9} K$.  Over the following few  
    tenth of a second, these layers are accelerated up to about 80,000 km/sec.  The further 
    evolution is governed by adiabatic expansion, modified by ongoing nuclear burning  and 
    radiative  cooling at the outermost layers.  After about 5 to 10 seconds, the expansion 
    is almost homologeous, i.e. the velocity is proportional to the distance.  Expansion 
    velocities exceed 30,000, 40,000 and 80,000 km/sec at the outermost $10^{-2}$, $10^{-4}$, and $10^{-5}$  $M_\odot$, 
    respectively.

The nuclear burning time scales increase with decreasing density.  As a consequence, 
only incomplete burning takes place throughout the outer half of the WD (in mass).  This 
property is well established by observations which show incomplete Si burning as well as 
explosive oxygen and carbon burning. Almost the entire WD undergoes burning with exception 
of the outermost $10^{-4} M_\odot$ (Fig. 2).  However, the shock front does not
stop its propagation and it still heats the surface layer.  In this context, we want to mention 
one of the major uncertainties related to the shock breakout.  The initial WD grows by 
accretion of H or He-rich matter.  As a result, we can expect He (and H)-rich surface 
layers in the outer few times $10^{-6} M_\odot$. Helium burns on signicantly shorter time scales 
compared to carbon and can produce additional nuclear energy even under low density conditions.
  The details depend sensitively on the amount of unburned H/He, mixing processes between the 
  H/He and the C/O layers. Test calculations showed variations in peak luminosity during the 
  outbreak by a factor of $\sim$3 from the specific model considered here.

\subsection{ The Shock Breakout}

%PAH
 The mechanism of the X- and gamma-ray production is similar to the Colgate
mechanism in core collapse supernovae where a shock steepens, however, with 
several distinguishing features.
In core-collapse SNe, the temperature increases by a strong detonation front whereas,
in thermonuclear supernovae and, within the now widely accepted delayed detonation
scenario, the front propagates even close to the surface as a weak detonation driven
by ongoing thermonuclear reactions which bring up the temperatures to 
billion degrees, and this front steepens close to the surface.
The $e^+,e^-$ pairs will be formed in the dense matter of the expanding envelope and annihilate almost instantly
so the compactness problem could be avoided and the high energy photons can escape the system.
 Most of the hard radiation
is emitted when the energy is released from dense matter when it becomes transparent due
to expansion.

As mentioned above, the outermost region is heated by the propagating shock front to a peak 
temperature of $\approx 1.5 \times 10^9$ K. The luminosity is governed by the decrease of the 
optical depth due to geometrical dilution, adiabatic cooling and, somewhat, energy production 
by ongoing nuclear burning.

The time scale for the burst luminosity is set by the rate of expansion and run time effects 
(Fig. 3). At the time of the outbreak, the radius of the object is about 9000 km, and
thermalization is almost instantaneous. As a result, the luminosity of the pre-expanded WD rises within about 1/30 of a 
second starting from a luminosity of about $10^{38}$ erg.  Subsequently,
 the matter undergoes  rapid acceleration from $\approx$10,000 km/sec to 80,000 km/sec on time 
 scales of a 0.2 to 0.3 seconds during which the radius increases by a factor of 7.  The result
  is a flash light curve with a fast rise and a slower decline which reaches a peak luminosity 
  of  $\approx 5 \times10^{49}$ erg/sec for a few hundredths of a second with an extended tail.

 The observable shock breakout is not thermal because multiple scattering
by thermal electrons (Pozdnyakov et al. 1976), the frequency dependence
of the opacity and, thus, different layers contribute to the spectrum, and
run time effects due to the extention of the source, i.e. the evolution 
over the expansion times.
 We have added a corresponding note in section 2.2

The monochromatic light curves show a low energy precursor in the 0.1 keV range, early hard 
radiation up to the MeV range followed by a rapid shift to hard X-rays on time scales of 0.1 
to 0.3 seconds. And a softening of the X-ray spectrum over 5 to 10 seconds. Note that the 
decreasing Compton opacity with wavelength and multiple scattering is crucial for hardening 
the radiation, an effect that is well known from hot stars.

\subsection{Interaction within the Progenitor System}
 
Up to now, we have considered the exploding WD in 'isolation'  neglecting the secondary  
X-rays and $\gamma $-rays in  context of thermonuclear explosions.  As discussed above, 
the WD is member of a close binary system with an accretion disk.  Likely, we have seen 
evidence for the interaction of the expanding envelope with its surroundings.  
 Potentially, this is a dominant contributor to the X-rays and, in particular, hard 
 $\gamma$-rays because we can directly tap into the kinetic energy of the outer layers 
 rather than the thermal reservoir. The available thermal energy is limited by the 
 nuclear energy production per nucleon ($\leq 3- 6$ MeV), whereas the available kinetic 
 energy of the outermost matter has gained kinetic energy originating from non-local 
 burning.  As a result, mean energies per atom are in excess of 100 MeV and about 8 \% 
 (i.e., $10^{50}$ erg) of the total explosion energy are deposited in the outer
  $10^{-2} M_\odot $ (Fig. 2).
 
We have evidence for this interaction that suggests that we can tap into the energy 
reservoir of the outer $10^{-3}-10^{-2} M_\odot $ and this is a common phenomenon.  
However, we have little information about the distribution and density of the 
surrounding matter which is critical for self-absorption, and the mechanism of 
transformation.  Is the process dominated by bremsstrahlung, thermalization, 
another process (e.g. magneto-hydrodynamical effects), or a combination of all?  
 From the specific energy of the particles in the ejecta and a H/He-rich surrounding,
%PAH
we may expect hard radiation somewhere between X-ray energies up to a few hundred keV.

Despite the uncertainties, we can estimate some of the properties along the lines of 
Sect. 2.  The total dimension of the  progenitor systems are of order 
$10^{11}$ cm and the accretion disk has an inner edge 
close to the exploding WD.  As a consequence, we can expect a rapid rise of the 
luminosity within a fraction of a second. Taking the expansion velocities from 
Fig. 2, the interaction will last less than a few seconds. Thereafter, the 
matter is swept up and will undergo adiabatic cooling.  As above, we must expect 
fast rise and slow decline light curves but on a longer time scale of several 
seconds with a comparable or slightly higher peak luminosity as compared to the shock breakout luminosity.  
Note that self absorption may severely reduce the observed fluxes or may somewhat increase the 
time scales because of intermittant trapping of energy. Because the observability 
depends sensitively on the geometry of the circumstellar matter and the orientation 
with respect to the observer, we must expect large individual variations.

\section{Correlating SNe Ia and GRBs}

	The goal of this section is to try to test the theoretical predictions that SN-Ia 
can give flashes of $\gamma$-rays that can trigger GRB detectors and look like GRBs
 with 1-10 second durations and fast-rise exponential-decay (FRED) light curves.  
 In particular, we will seek to find these SN-Ia Flashes by looking for correlations 
 between catalogued Type Ia SNe and catalogued GRBs.  The result can then be used to 
 place limits on the luminosity of SN-Ia Flashes.
	
	The procedure is to compile a list of all known SN-Ia events (with an upper limit 
on their distance), estimate their date of core collapse, and seek a cataloged GRB 
with a consistent date and position on the sky.  The reason to go in this direction 
is that we know the SN-Ia collapse occured at a specific time and direction, so we 
have the simple question of asking whether any associated SN-Ia Flash was detected 
by a satellite GRB detector.  In general, we cannot know the exact time of the collapse, 
so we cannot know whether any particular GRB detector was pointed in the right direction 
at the right time.  This makes our test a statistical one.  We can know the number of 
SN-Ia events that might have been covered by GRB experiments, and we can calculate what 
fraction of the known SN-Ia events are likely to have good coverage, so we can estimate 
how many of these known SN-Ia events should appear in GRB catalogs if the SN-Ia Flashes 
are brighter than the detection limit.  If many Flashes are expected but none are detected, 
then we will have a limit on the Flash brightnesses.  When combined with the upper limit on 
the distances, this will translate into an upper limit on the Flash luminosity.  And this 
can then be compared with our earlier theoretical predictions. 

	This study is possible only because SN-Ia Flashes have properties similar 
to GRBs.  In particular, SN-Ia Flashes have typical peak temperatures corresponding 
to ~100 keV which cools substantially throughout the event, and light 
curves with fast rises and roughly exponential declines with time scales 
of a few seconds.  This description of a SN-Ia Flash is identical to those of the 
multitude of FRED bursts.  If a SN-Ia Flash is bright enough, then it would 
be 'hiding' inside the GRB catalog as an apparently ordinary FRED burst.

	The FRED shape is a common light curve shape for individual 
pulses within a burst.  Yet bursts with multiple FRED pulses cannot be a 
SN-Ia Flash since only one white dwarf can collapse to make only one FRED.  We do 
not know how much fluctuations to expect from turbulence in the outer 
layer of a WD or any surrounding gas, so the basic FRED shape can well have superposed 
spikes or modulations of perhaps large amplitude.  A perusal of the light curves 
displayed in the first BATSE catalog (Fishman et al. 1994) shows that 
roughly 50\% of all BATSE triggers are single FREDs (perhaps with 
significant fluctuations around the basic FRED shape) with time scales 
from 1-10 seconds.  

\subsection{GRBs Included}

	For this study, we will use GRBs from three satellite experiements; BATSE on the
 {\it Compton Gamma-Ray Observatory}, HETE, and Swift.

\subsubsection{BATSE Bursts}

	The BATSE detectors covered the entire visible sky for 9.1 years, and this provides 
a large coverage with deep limits.  We have adopted the BATSE 4B burst catalog 
(Paciesas et al. 1999) as well as its extensions up until the date of the satellite 
re-entry (BATSE GRB Team 2001).  This covers 2702 triggered GRBs from 19 April 1991 
until 26 May 2000 (an interval of 3323 days).  This is an average of 0.813 triggered
 bursts per day.  After accounting for Earth-blockage, SAA passages, and other 
 inefficiencies, BATSE covered the entire sky for this time interval with an average 
 efficiency of 39\% (Fishman et al. 1994; Paciesas et al. 1999).

	When we seek positional coincidences between precisely-known SN positions and the BATSE 
positions, we must estimate the uncertainties in the BATSE positions to know whether the 
positions are coincident.  We will adopt the two-sigma positional error radius as a 
reasonable compromise between missing true connections by making the radius too small 
and adding false connections by making the radius too large.  (Repeated calculations 
with 1.0 and 3.0 sigma radii yield the same results as below except with larger 
uncertainties.)  The cataloged one-sigma radius for each individual burst is for the 
statistical error only, and must be increased by the systematic error of 2.0 degrees 
(Briggs et al. 1999) added in quadrature so as to get the total radius of the one-sigma 
positional error circle ($\sigma_{BATSE}$).  The two-sigma radius is just twice the 
one-sigma radius.  For the purposes of this paper, the error regions will be assumed 
to be circular in shape, even though individual bursts will have moderate distortions 
from this ideal.

	The sum of the areas for the two-sigma error regions over all 2702 bursts is a total 
of 1.16 million square degrees.  The average burst area is 431 square degrees, which 
is 1.05\% of the sky.

	The BATSE catalogs are ideal for estimating the number of bursts with a peak flux 
brighter than some stated threshold.  For example, the cumulative distribution of 
bursts brighter than some give peak flux in the 50-300 keV energy band over a one 
second interval is given in Figure 6c of Paciesas et al. (1999).  To convert this 
number to a rate (with units of bursts per year) for the whole sky with perfect 
efficiency, we have to divide by the efficiency (39\%) and divide by the number of 
years in the 4B catalog (5.37 years).  So, for a peak flux of 1.0 photons cm$^{-2}$ s$^{-1}$ 
or brighter, there are 400 BATSE bursts in the 4B catalog, and this gives a total 
count of 190 bursts per year appearing over the entire sky.  %To a limiting peak 
flux of 0.5 photons cm$^{-2}$ s$^{-1}$, there are 700 bursts in the 4B catalog, which 
translates into 330 bursts per year over the entire sky.

\subsubsection{HETE Bursts}

	HETE has detected many GRBs from roughly 2001 to 2005, while providing fast information 
to the ground so as to allow rapid follow-up of burst positions.  We will use a catalog 
of 69 HETE bursts with accurate positions detected in the 4.0 year interval from 2001.0 
to 2005.0 (HETE Team 2006; see also Greiner 2006).  

	Most HETE bursts have received substantial ground-based follow-up optical imaging in 
the hours and days after the burst.  Greiner (2006) presents an extensive compilation 
of reports from this vast program.  This follow-up work would likely have discovered 
any nearby SN-Ia event at the position of the GRB.  Thus, there is no real expectation 
that a comparison of nearby SN-Ia lists will turn up any GRB/SN connections.  As such, 
we are merely trying to place limits on the brightness and luminosity of any SN-Ia flashes
that arise from known Type Ia supernovae.

	The HETE bursts all have arc-minute positions, while the supernovae will all have 
arc-second positions.  Should any HETE position contain any SN position, we can be 
confident that the positional overlap is not due to any random coincidence, and thus 
we would conclude that the GRB and SN are causally related.  So, unlike for BATSE, we 
don't have to worry about false alarms due to coincidences.  Rather, the primary 
question will be with whether HETE was actually looking at the SN position at the 
time of the shock breakout.
	
	Sakamoto et al. (2005) present a compilation of the peak fluxes for many HETE bursts.
  We have used the peak fluxes from 30-400 keV over one-second time intervals to 
  construct a brightness distribution.  This distribution should be rising as a power 
  law to low peak fluxes yet with a break due to the HETE threshold.  In the observed 
  distribution, we see that the 1-3 photons cm$^{-2}$ s$^{-1}$ bin already is 
  substantially lower than expected from the numbers in the 3-9 photons cm$^{-2}$ s$^{-1}$
   bin, hence suggesting a break at around 2 photons cm$^{-2}$ s$^{-1}$.  But HETE detected
    many bursts going to peak fluxes of 0.1 photons cm$^{-2}$ s$^{-1}$ and fainter.  This 
    broad threshold will have a 50\% detection probability around 1 photons cm$^{-2}$ s$^{-1}$.
      This detection threshold corresponds to roughly 1.7 $\times 10^{-7}$ erg cm$^{-2}$ s$^{-1}$.
        From the previously stated BATSE result, this limit corresponds to 190 GRBs per 
        year over the whole sky.  As the HETE catalog reports on 69 bursts per 4.0 years,
         the overall fractional sky coverage by HETE must be roughly 9\%.  That is, any 
         particular SN-Ia from 2001 to 2005 will have an average chance of 9\% that HETE
          was looking at its Flash.  The uncertainty on this fraction will primarily be 
          in the systematics of the comparison between the satellites, which we estimate
           to be $\pm$3\%.
	
\subsubsection{Swift Bursts}

	Swift (Gehrels et al. 2004) was launched in late 2004 and is still recording bursts at 
a fast rate.  We will use the Swift catalog for the 1.75 year interval from 2005.0 to 
2006.75 (Swift Team 2006; see also Greiner 2006).  This interval has 140 Swift GRBs 
with accurate position and comprehensive optical follow-up imaging.  As with HETE, any 
nearby SN-Ia associated with the event would almost certainly have been quickly recognized
 (so we are not expecting to find any such connections) and the Swift positions are 
 arc-minute in size (so there is no real chance of random coincidence producing a false alarm).
	
	The Swift burst detector has a half-coded field of view of 1.4 steradians (Gehrels et 
al. 2004), which is 11\% of the sky.  In practice, roughly 40\% of the detected Swift 
GRBs are with less than half-coding, so the effective coverage of the sky is 19\%.  The 
operations of the spacecraft keep the field of view of the burst detector outside of 
Earth occultation.  With down-time largely being the small fraction due to the SAA, 
we estimate that the Swift sky coverage is roughly 18\%.  We estimate the uncertainty 
in this fraction to be $\pm$6\%.
	
	The Swift web page tabulates the peak fluxes from 15-150 keV over one-second time 
intervals.  We have constructed a brightness distribution for these bursts.  We see 
a fairly sharp threshold, with the 1-3 photons cm$^{-2}$ s$^{-1}$ bin having the 
numbers expected (based on a power law extrapolated from the numbers in the brighter 
bins), while the 0.3-1 photons cm$^{-2}$ s$^{-1}$ bin is down by almost a factor of 
four.  On this basis, we take the Swift threshold to be close to 0.7 photons cm$^{-2}$ s$^{-1}$. 
 This threshold is for a passband of 15-150 keV, and this is equivalent to a threshold 
 of roughly 0.5 photons cm$^{-2}$ s$^{-1}$ for a pass band of 50-300 keV.  From BATSE, 
 we expect that there should be 330 bursts per year over the whole sky to this threshold. 
  With Swift seeing 140 bursts per 1.75 years, this implies a fractional sky 
  coverage of 24\%.

\subsection{Supernovae Included}

	For this study, we have adopted the Asiago Supernova Catalog 
(Barbon et al. 1999), which can be obtained as an up-to-date version 
on-line.  From this catalog, we have extracted supernovae that are explicitly
 identified as being of Type Ia with dates of explosions between the start 
 and stop dates for each of our GRB data sets.  We will primarily be looking
  at those SNe whose host galaxy has a radial velocity (RV) of 3000 km s$^{-1}$ 
  or less (corresponding to a distance of 43 Mpc for a Hubble constant of 70 km s$^{-1}$ Mpc$^{-1}$).  
  The reason for this cut is since we 
expect the SN-Ia Flashes to be primarily visible from the nearest SNe, while the 
inclusion of more distant events will only dilute the statistics.  We have tried 
varying this RV cutoff, yet we reach the same conclusions.  

\subsubsection{For the BATSE catalog}

	The BATSE catalog covers from 19 April 1991 to 26 May 2000.  Our sample of Type 
Ia supernovae in the Asiago catalog with $RV < 3000$ km s$^{-1}$ includes 38 SNe.
  We have chosen to eliminate 5 events (SN1991ak, SN1993Z, SN1993af, SN1994aa, and 
  SN1998cn) from this sample based on a criterion that the day of explosion must 
  have an uncertainty of 10 days or better.  (Again, the relaxation of this criterion 
  does not change our results.)  Thus, our primary sample of nearby Type Ia SNe 
  contains 33 events (see Table 1).

	For each supernova on this list, we determined the date of peak 
brightness ($T_{peak}$) from the literature.  These dates are presented in Table 1.  
Each date also has an estimated one-sigma uncertainty.  Some of these 
uncertainties are from formal fits of the light curve to templates, and 
these are assigned an uncertainty of $\pm$2 days.  However, if a formal 
template fit was made to data starting after five days of the claimed 
maximum light, then the derived date is assigned an uncertainty of $\pm$5 
days.  The 'snapshot' method of Riess et al. (1998) is regarded as having 
an uncertainty of $\pm$7 days.  If a peak date is assigned based on a 
spectrum being 'near maximum', then an uncertainty of $\pm$7 days was 
assigned.  Other cases have uncertainties as tabulated.

	For each supernova, we determined the offset between the date of 
the explosion and the date of the peak in the B-band.  We have adopted an offset 
of 19.8 days for a supernova (Riess et al. 1999b) with the stretch factor (S) 
equal to unity.  The offset for other stretch factors is $19.8/S$ days.  The 
measured quantity $\Delta m_{15}$(B) is taken as $1.96(1/S-1)+1.07$ (Krisciunas et 
al. 2000).  We have adopted an uncertainty of $\pm$2 days in the offset for 
SNe with a measured decline rate.  If the SN does not have a measured 
decline rate, we adopt $S=1$ and an uncertainty of $\pm$4 days for the offset.  
If $S<0.75$, we take the uncertainty in the offset to be $\pm$4 days.

	The date of the explosion (i.e., when the collapse starts the 
rapid brightening in the light curve) is found by subtracting the offset 
from the date of the peak in brightness.  For each SN, we compared this 
with the first reported positive detection of the SN, and fortunately we 
had no conflicts.  The result is a date of explosion ($T_{exp}$) for each of the
 33 SNe in Table 1.

	As a null test for the significance of any correlation, we have 
also constructed a set of dates for each SN where the offset is {\it added} to 
the date of peak.  This produces a set of dates (that certainly does 
not include the true dates of the explosions) which have similar 
distributions as the real dates of explosions.  The idea will be that the 
number of GRB/SN matches for this 'time-reversed' set should provide a 
measure of the false alarm rate for matches.

	We have also created a sample of SN-Ia events with a radial velocity between 
5,000-10,000 km s$^{-1}$.  The events in this high RV sample 
typically have $\sim$4 times the distance as SNe in our primary sample ($<$3,000 
km s$^{-1}$).  Thus, any SN-Ia Flashes in the high velocity sample will be $\sim$16
 times fainter than in our primary sample.  Provided that SN-Ia Flashes do not have a 
broad luminosity function, the presence of a substantial number of SN-Ia Flashes in 
the high velocity sample would force the presence of many and bright SN-Ia Flashes 
in the low velocity sample.  This would be obvious, whereas the few 
possible matches involve faint bursts near the BATSE threshold.  As such, 
any SN-Ia Flashes from the distances corresponding to 5,000-10,000 km s$^{-1}$ would 
have been below the BATSE threshold.  One implication of this is that the high 
velocity sample can be used as a control sample for estimation of the 
number of coincidental matches in our primary sample.

\subsubsection{For the HETE and Swift Catalogs}

	The effective dates for the HETE and Swift GRB data sets are 2001.0 to 2005.0 
and 2005.0 to 2006.75 respectively.  The Asiago SN catalog returns 30 and 16 
SN-Ia events (with $RV < 3000$ km s$^{-1}$), respectively.  These SNe are not 
separately listed because there are no SN/GRB matches.

\subsection{The Observed Matches}

	If some (or all) Type Ia events produce an observable burst of 
gamma radiation (a SN-Ia Flash), then there should often be a matching burst 
detected by the satellites.  How many of the nearby SN-Ia in the Asiago catalog 
have a match with a cataloged $\gamma$-ray event?

\subsubsection{BATSE/SN Matches}  

	For each of the 33 SNe in Table 1, the BATSE catalog was searched for any event 
that coincided in both time and direction.  That is, the dates of the GRB and the
 SN explosion must match to within the stated error bars and the two positions 
 must agree to within the two-sigma positional error bar.  For all  the matches,
  Table 1 lists the BATSE burst, the angular separation between SN and GRB ($\Theta$)
   divided by $\sigma_{BATSE}$, and the time difference between the GRB and SN 
   explosion ($\Delta T$).  In italics, we have also added the information for 
   the matches where the angular separation is from 2 to 3 sigma.

	We checked to see if the InterPlanetary Network (IPN; Hurley et al. 1999; 2000)
 could be used to reduce the size of the GRB error boxes.  The idea is that 
 smaller error boxes will either increase the confidence in the GRB/SN connection 
 or will eliminate the possibility.  Unfortunately, none of the listed bursts have
  IPN error boxes.

	In all, for the 33 nearby Type Ia SNe during the BATSE era, four 
GRBs are consistent in time and within the two-sigma positional error box.  
If we expand the positional error boxes for the GRBs to a three-sigma cut, 
then we have 10 matches.  If we restrict the cut to one-sigma boxes, then 
we have two matches.

	The individual bursts with good matches are good candidates for 
SN-Ia Flashes.  As such, the associated gamma-ray events were examined closely to 
see if they shared any common features.  For example, we looked at the 
light curve shape, the duration, the smoothness of the light curve, the 
spectral lag of the light curve, and the BATSE hardness ratio.  We could 
find no common traits shared by the matches.  However, we found that all 
the matches were with faint BATSE bursts (which must necessarily have 
large positional error boxes).  This is exactly what we would expect if 
the matches are random coincidences.

	Two of our matches (SN1991bg/GRB911125 and SN1997bq/GRB970331) are 
events with a FRED light curve and time scales from 1-10 seconds.  As 
such, these two events are particularly interesting as possible SN-Ia Flash 
candidates.  However, roughly half of all GRBs appear as single-episode 
fast-rising events with a slower decline (perhaps with fluctuations) with 
time scales of 1-10 seconds.  So the presence of two FREDs amongst our 
four matches implies that the FRED shape of two matches is not 
significant.

	As a statistical control, we also checked a group of distant SNe 
for matches with GRBs.  In particular, we examined 66 Type Ia events with 
host galaxy radial velocities from 5,000 to 10,000 km/s as extracted from 
the Asiago catalog.  We searched for matches to GRBs which occurred 
anytime in the month prior to the date recorded in the Asiago catalog 
(usually the discovery date).  This set of date ranges and positions 
should have similar distributions as for our primary set of nearby SNe.  
The result was 17 matches within the two-sigma GRB error radii.

	For the 'time-reversed' set of explosion times, a similar 
collection of matches was made.  The number of matches was 1, 4, and 9 for 
the one-sigma, two-sigma, and three-sigma error boxes respectively.

\subsubsection{HETE/SN and Swift/SN Matches}

	We examined our lists of selected SN-Ia events against the HETE and Swift 
GRB lists.  No matches were found, with all pairs being far from acceptable.
  This is not surprising, as any SN that would have appeared in a GRB error 
  box would likely have been quickly discovered and widely known by now.

\subsection{The Expected Number of Coincidental BATSE/SN Matches}

	Our basic sample of 33 SNe has resulted in 4 matches (for two-sigma error 
boxes) with BATSE bursts.  Some number (perhaps all) of these matches could
 be due to random coincidences.  This section will evaluate the number of 
 expected coincidences by three methods.

	Each SN in Table 1 has a stated uncertainty for the date of the 
explosion.  As this interval is two-sided, the time during which a GRB 
coincidence would be accepted is twice that amount.  When summed over all 
33 SNe, the total time interval is 364 days.  The probability that a GRB 
will randomly occur with an error box that includes the SN position and at 
a time consistent with a specified SN is simply the average burst rate 
(0.813 burst/day) times the fraction of the sky that is covered by an 
average GRB error circle (1.05\%) times the total time interval in which 
the explosion is constrained (364 days).  Thus, the expected number of 
coincidental matches is 3.1 bursts.

	This simple calculation will be inexact if the GRBs and SNe are 
clumped in time and position.  Indeed, the GRBs were discovered at rates 
which vary with the BATSE trigger criteria and solar activity while the 
SNe rates peak in the seasons when high galactic latitudes are near the 
meridian at midnight.  Also, the SNe are strongly concentrated towards the 
galactic poles while the BATSE bursts are essentially isotropic.  These 
nonuniformities will create correlations that will systematically distort 
the estimated number of coincidental matches by some small percentage.

	The set of 'time-reversed' dates provides a sample with similar 
distributions in time and direction as the real sample of explosion dates.  
The number of matches between this time-reversed data set and the real 
BATSE burst catalog will share all the effects as caused by the various 
nonuniformities, yet none of the matches can be caused by a real SN-Ia Flash.  
Thus, the observed number of time-reversed matches should be equal (within 
the usual Poisson statistics) to the number of coincidental matches 
contained in Table 1.  This number has already been identified as 4 
matches in the time-reversed data set.  The uncertainty is $\pm$2 bursts.

	The set of SNe with radial velocities between 5,000 and 10,000 
km/s has the same distribution over time and over the sky as our primary 
set with radial velocities of $<$3,000 km/s.  Thus, the rate of coincidental 
matches between this high velocity SNe sample and BATSE GRBs should be the 
same as for our primary low velocity sample.  Out of 66 high velocity SNe, 
17 had matches with the BATSE catalog.  Each high velocity SNe had a time 
interval of one calendar month (an average of 30.4 days) for the match, 
with a total time interval of 2006 days.  Thus, the control sample had an 
average of one match every 118 days.  The 33 SNe in our primary sample had 
a total time interval of 364 days, so we would expect 3.1 matches by 
coincidence alone.  The uncertainty in this estimate comes by propagating 
the Poisson noise for the original 17 count, so that the error bar will be 
24\%.  In all, this control sample gives a rate of 3.1$\pm$0.7 coincidental 
matches for our primary sample.

	Our primary sample of 33 SNe produced 4 matches out of which some number are 
purely coincidental matches.  We have just estimated the number of 
coincidental matches to be 3.1, 4$\pm$2, and 3.1$\pm$0.7.  The first estimate 
suffers from unknown systematic errors due to the clumpiness of GRBs and 
SNe in time and space, so we will only use it to provide confidence that 
the more complex methods are not greatly in error.  The remaining two 
estimates are independent and hence can be combined as a weighted average 
to yield our final value for the false alarm rate of 3.2$\pm$0.7 coincidental 
matches.  When the number of coincidental matches is subtracted from the 
4 observed matches, we are left with a 0.8$\pm$0.7 matches as being of 
non-coincidental origin.  This is consistent with a null detection of SN-Ia 
Flashes.  That is, the observed number of matches is consistent with being 
entirely due to random overlaps with no physical connection between the gamma ray event and the SNe.  Alternatively, we can place a one-sigma upper limit on the number of SN-Ia Flashes in the BATSE data as 1.5 matches.

\subsection{Limits on Luminosity and Covering Fraction}

	Our searches for BATSE, HETE, and Swift are all consistent with zero SN/GRB 
matches.  For BATSE, the large error boxes and the resultant likelihood of 
chance coincidences only allows us to place a one-sigma upper limit of 1.5 
matches.  For HETE and Swift, with their small error boxes, we know that 
there are 0 and 0 matches respectively.  Our lack of matches could be due 
to the detectors not happening to point at the SN-Ia at the time of collapse, 
the SN-Ia luminosity being too low to allow for detection, or the shock breakout 
being covered by some object (the companion star or the accretion disk) in the 
progenitor.  With the above information, we can calculate how many of the nearby 
SN-Ia events will likely have been in the field-of-view of each of the satellites. 
 From this, we can then place a limit on the luminosity of the SN-Ia Flashes.  
 Alternatively, we could place limits on the covering fraction, where the companion 
 star or the accretion disk hides the shock breakout from view here at Earth.

\subsubsection{BATSE Limits}

	BATSE had a significant fraction of deadtime in its coverage of any one location on
 the sky.  This fraction is 39\% (e.g., Fishman et al. 1994), and represents normal 
blockage by the Earth, deadtime due to the South Atlantic Anomaly, and 
other inefficiencies.  This means that out of the 33 SNe in our primary 
sample, only 12.9$\pm$3.6 SN-Ia Flashes {\it could} have been detected by BATSE.  The 
difference between 12.9$\pm$3.6 and 0.8$\pm$0.7 could be either due to some 
fraction of the SN-Ia Flashes being covered (say, by an accretion disk) that hides 
the flash from some directions or by the luminosity of SN-Ia Flashes being so low 
that many of the SN-Ia Flashes in the sample volume (out to 43 Mpc) would be below 
the BATSE threshold.

	BATSE was operating and monitoring the sky at the times of the 
shock breakouts for 12.9$\pm$3.6 nearby Type Ia SNe.  The lack of observed 
GRB/SN matches (above chance coincidence) could be simply due to the SN-Ia Flashes 
being too faint for BATSE to detect.  The BATSE trigger threshold is 
$3 \times 10^{-8}$ erg s$^{-1}$ cm$^{-2}$ or 0.2 photon s$^{-1}$ cm$^{-2}$
 for the 1.024 second trigger time scale (Fishman et al. 1994).  The 12.9 nearby 
 supernova must be at a distance closer than 43 Mpc.  For them not to be detected 
 due to low luminosity, their luminosities must all be below 
 $7 \times 10^{45}$ erg s$^{-1}$ (in the 50-300 keV energy band for the usual 
 BATSE trigger).  If the average SN-Ia Flash luminosity is lower than this upper 
 limit, then BATSE will detect events only within a smaller volume and this will 
 reduce the expected number of matches.  To reduce the expected number from 12.9 
 (for relatively high luminosity SN-Ia Flashes) to 1.5 (the one-sigma upper limit 
 on the observed matches), the radius of the volume must decrease from 43 Mpc to 21 
 Mpc.  A detection limit of 21 Mpc corresponds to a limit on the average SN-Ia Flash
  luminosity of $ 2 \times 10^{45}$ erg s$^{-1}$ (for the 50-300 keV band).  For 
  a spectrum with a spectral peak at 100 keV, the bolometric peak flux will be 2.4 
  times the 50-300 keV peak flux.  Thus, if the lack of observed BATSE/SN matches 
  is caused by SN-Ia Flashes having low luminosity, then our observational limit is 
  $5 \times 10^{45}$ erg s$^{-1}$ for the bolometric peak flux.
  
  	The limit derived in the previous paragraph applies to the BATSE trigger time scale of 1.024 seconds, as is optimal for the flash arising from interaction with the accretion disk.  But the shock breakout has a time scale of peak emission of roughly 0.1 sec, so the flux averaged over any 1.024 second interval would be smaller by a factor of ten.  For these fast time scales, BATSE also has triggers that operate on 0.064 and 0.256 sec time scales.  The BATSE trigger threshold is $1.5 \times 10^{-7}$ erg s$^{-1}$ cm$^{-2}$ or 1.0 photon s$^{-1}$ cm$^{-2}$
for the 0.064 second trigger time scale (Fishman et al. 1994).  For a detection limit of 21 Mpc, the limit on the Flash luminosity is $ 1 \times 10^{46}$ erg s$^{-1}$ for the 50-300 keV band or $2 \times 10^{46}$ erg s$^{-1}$ for the bolometric peak flux.

	If some fraction, F, of the SN-Ia Flashes are uncovered and visible from 
afar, then the expected number of SN-Ia Flashes in our sample would be  
$12.9 \times F$.  (This assumes that all SN-Ia Flashes are luminous enough to be 
detected by BATSE out to distances of 43 Mpc.)  The best estimate of F is then 
0.8/12.9=0.062.  However, the uncertainty in F is large.  For a value of 
F=0.12, the expected number of matches will equal the one-sigma upper 
bound on the number of observed SN/GRB matches.  Thus, the near zero number 
of matches might imply that $>88$\% of the SN-Ia Flashes are usually covered by 
something like an accretion disk and are hidden from view.

\subsubsection{HETE Limits}

	HETE was up and watching the sky during the time when 30 nearby SN-Ia events occurred.
  Of these 30, we expect that 9\%$\pm$3\% will have HETE pointed in the right direction
   at the right time.  So we expect 2.7$\pm$0.9 SN-Ia Flashes that HETE could have detected.
     But HETE saw 0 matches.  For a Poisson distribution, the probability of seeing zero
      matches if 2.7 matches are expected is 7\%.  Thus, at a little less than a two-sigma
       confidence level, we can account for the lack of HETE/SN matches as being due to the
        randomness of HETE pointing.  As such, we realize that any limits from HETE will be
         weak.
	
	Nevertheless, taken at face value, HETE likely {\it was} looking at the correct time and 
direction to see several SN-Ia flashes - but saw nothing.  The HETE threshold is roughly 
1.7 $\times 10^{-7}$ erg cm$^{-2}$ s$^{-1}$ with SN distances no farther than 43 Mpc.  This 
forces the SN-Ia Flashes to have a luminosity of less than $4 \times 10^{46}$ erg s$^{-1}$. 
 With a bolometric correction, the HETE limit on the total peak luminosity will be 
 $7 \times 10^{46}$ erg s$^{-1}$ if the lack of matches is due to the faint luminosity 
 of the SN-Ia Flashes.
		
	Alternatively, we could require a covering fraction that would reduce the expected 2.7
 matches down to some smaller number, but this covering fraction could be quite small and 
 we'd still have an acceptable case with zero observed matches.

\subsubsection{Swift Limits}

	Swift was watching when 16 nearby SN-Ia bursts were visible.  We expect that 18\%$\pm$6\%
 will have Swift covering the shock breakout.  Thus, 2.9$\pm$1.0 SN-Ia Flashes are expected 
 to be detected by Swift if they were bright enough.  Swift saw 0 matches, and this indicates 
 that the SN-Ia Flashes were too faint either due to low luminosity or coverage within the 
 progenitor system.  With almost identical statistics as in the HETE case, we realize that 
 there is a roughly two-sigma chance that the zero matches is simply due to poor luck in sky coverage.
	
	Nevertheless, Swift was likely looking at the correct time and direction to see several SN-Ia flashes 
- but did not trigger on any flash.  The Swift threshold is roughly 1 $\times 10^{-7}$ erg cm$^{-2}$ s$^{-1}$ 
with SN distances no farther than 43 Mpc.  Then, the SN-Ia Flashes must have a luminosity of less than 
$2 \times 10^{46}$ erg s$^{-1}$ in the 15-150 keV band.  The Swift limit on the bolometric peak luminosity 
will be $5 \times 10^{46}$ erg s$^{-1}$.
		
	Again, a covering fraction could reduce the expected 2.9 matches down to some smaller number.  But any 
limits on the covering fraction are weak.
	
\section{Discussion}

	Our theoretical and observational conclusions on the peak brightness are in contradiction. 
That is, we predict that SN-Ia
Flashes will appear as ordinary FRED GRBs with peak luminosities of $\sim 10^{48}-10^{50}$ erg s$^{-1}$, however, our observational constraints show SN-Ia Flashes to have bolometric peak luminosities of less than
$\sim 10^{46}$ erg s$^{-1}$.  The observational limits are strong since many of the known nearby 
SN-Ia collapses must have been observed by many GRB satellites, and so our peak luminosity limits are 
robust.  And theoretically, there inevitably must be some sort of Flash caused both by the inevitable 
shock breakout and the inevitable collision of the ejecta with the accretion disk.  
	
	One possible resolution of this discrepancy is that the shock breakouts are usually covered up from Earth
view, perhaps by the inner edge of an accretion disk that is much thicker than a WD radius so as to 
shield the WD and its shock breakout from most of the sky.  Within this idea, the $\gamma$-ray flash 
from the ejecta ramming into the disk itself would also be absorbed by the outer parts of the disk 
itself.  Within this idea, there would likely be two funnel regions around the poles of the accretion 
disk that are largely clear, such that an observer in a polar direction could still see the SN-Ia 
Flash.  From our data, with the three satellites on patrol for $\sim$15 known nearby SN-Ia events, 
the clear sky fraction would have to be less than 7\% or so.  This would give a half-opening angle 
of the funnel to be less than 20$\degr$ or so.
	
	Finally, we would like to point out the importance of SN-Ia Flashes (should one be detected) as 
diagnostical tools to constrain explosion physics and the progenitor system.  There is growing 
evidence that mergers of pulsating delayed detonations contribute to the population of SNe~Ia 
(H\"oflich \& Khokhlov 1996, Quimby et al. 2005).  Even within the delayed detonation scenarion and as mentioned in Section 2,
the luminosity of the shock breakout depends on the chemistry of the outermost layers. In principle, 
observations may allow us to learn about the accretion history and possible mixing processes. 
Spherical symmetry implies that the nuclear outbreak occurs simultaneously. Thus, the rise to 
maximum light is minimized.  On the other hand, recent observations of late time spectra 
(H\"oflich et al. 2004) and early time polarization (Wang et al. 2006, H\"oflich et al. 2005) suggest off-center DDTs which
imply run time effects of $\approx $ 0.3 to 0.4 seconds. In principle, direct measurments of 
the rise (and the resulting change in the time evolution of the  spectral energy distribution) 
to maximum may allow to measure these run-time effects.  Finally, we have presented estimates 
for the burst properties but without detailed calculations.  These calculations would require 
full 2D or 3D calculations for the interaction of the envelope with the surrounding disk, and 
detailed studies of the radiation processes involved.  Unfortunately, the initial conditions 
of the disk are not well known but a knowledge is critical for detailed predictions. Alternatively, 
a more detailed analysis of individual events may help to constrain the properties which may include 
both  the low and high energy photons. For the future, we plan to address gradually all these aspects.

\section{Conclusions}

	(1)  The shock breakout from the white dwarf in a Type Ia SN and the interaction of the rapidly 
expanding envelope with matter within the progenitor system  will produce a burst of X- and  $\gamma$-
radiation (SN-Ia Flashes). For the shock breakout, typical peak temperatures are  $\sim $ 100 keV at 
peak with durations of roughly 0.1 to 0.3 seconds and peak luminosities of $\sim 10^{49}-10^{50}$ erg s$^{-1}$.
For the envelope smashing into the accretion disk, the expected durations will be roughly 1-10 seconds 
and the characteristic photon energies are expected to the MeV range while the peak luminosities are
expected to be of order $10^{48}-10^{50}$ erg s$^{-1}$.  Both mechanisms will produce fast-rise
 exponential-decay shapes in their light curve.  The temperature of the fireball should substantially 
 cool over the duration of the burst.

	(2)  Our predicted SN-Ia Flashes events should look similar to the FRED GRBs, and might already be 
in the GRB catalogs.  We have looked for matches between the Asiago SN catalog and the BATSE, HETE, 
and Swift burst lists.  The BATSE constraints are the most decisive and restrictive, primarily because 
its sky coverage (in units of year-steradians) was a factor of ten times larger than the other two 
satellites.  For BATSE, we have identified 33 Type Ia SNe from 19 April 1991 and 26 May 2000 whose 
host galaxies have radial velocities of $<$3,000 km s$^{-1}$ (i.e., nearer than 43 Mpc for a Hubble 
constant of 70 km s$^{-1}$ Mpc$^{-1}$) and whose date of explosion can be determined to within 10 
days.  Four BATSE bursts were found to have been consistent in position (within the two-sigma BATSE 
circle) and time.  The number of chance coincidence was determined to be 3.2$\pm$0.7, so the observed 
number of matches due to SN-Ia Flashes is 0.8$\pm$0.7.  This non-detection of SN-Ia Flashes can be 
used to either limit the fraction of directions not covered (e.g. by an accretion disk) to be $\leq 12 $ \%
or to limit the average bolometric peak luminosity of $\sim 10^{46}$ erg s$^{-1}$.  With the HETE 
and Swift satellites, we expect to see a total of 5.6$\pm$1.3 SN-Ia Flashes from {\it known} nearby SN-Ia 
collapses, whereas zero were seen.  This limits the peak luminosity of SN-Ia Flashes to be less than 
$5 \times 10^{46}$ erg s$^{-1}$.

	(3)  The expected and observed peak luminosities are inconsistent by several orders of magnitude.
  The GRB satellites were watching for SN-Ia Flashes from a total of something like fifteen known events 
 yet detected nothing, while the shock breakout and ejecta/disk interaction also are inevitable. 
  We do not think that either the observational or theoretical results can be wrong by several 
  orders-of magnitude. 
   We suggest that the initial flash of the shock breakouts and the onset of the interactions with 
 accretion disk are hidden. Because the initial flash is not seen either, it is unlikely that material
originates from a thin accretion disk or the donor star which is redistributed during hydrodynamical
interactions during the explosion (e.g. Marietta et al. 2000) but it suggests  a thick accretion
disk or common envelope that nearly smothers the WD. We note that this result is also consistent
with the high covering factor needed to correct for the discrepancies in numbers  between observed and expected
supersoft X-ray sources which are regarded as possible progenitors (Rappaport et al. 1994, Kahabka \& van den Heuvel 1997).
	
	(4)  Even one detection of a SN-Ia Flash will tell a substantial amount about 
the physics of Type Ia SNe as well as about the composition (and hence recent accretion history) of 
the outer layers as a guide to the progenitor type.

 Finally, we also would like to stress the limits. Though multidimension effects will hardly effect the
order of magnitude and the conclusions, multi-dimensional effects will become important. Secondly,
the current data set is very limited and allow to estimate a covering factor of more than 88\% but,
to e.g. distinguish high scale heights of accretion disks from common envelopes demands a better statistics
and at least one positive detection with good time and frequency coverage.

\acknowledgements
This work was supported by the grants from  NSF (AST0307312  to H\"oflich) and from NASA (NAG5-7937 to H\"oflich and NNG06GH07G to Schaefer).

\clearpage

\begin{deluxetable}{llllllll}
\tabletypesize{\scriptsize}
\tablecaption{SN-Ia Matches with BATSE GRBs 
\label{tbl1}}
\tablewidth{0pt}
\tablehead{
\colhead{SN}   &
\colhead{RV (km/s)}   &
\colhead{Ref.\tablenotemark{a}}   &
\colhead{$T_{peak}$}  &
\colhead{$T_{exp}$}  &
\colhead{GRB}  &
\colhead{$\Theta / \sigma _{BATSE}$}  &
\colhead{$\Delta T$ (days)}  
}
\startdata
1991 T	&	1732	&	1	&	Apr 28.7$\pm$2(B)	&	Apr 8$\pm$3	&	...	&	...	&	...	\\
1991 X	&	2626	&	2	&	May 5$\pm$7	&	Apr 15$\pm$8	&	910423	&	0.57	&	8$\pm$8	\\
1991 bg	&	913	&	 3-5	&	Dec 13$\pm$2(B)	&	Nov 29$\pm$5	&	911125	&	0.6	&	4$\pm$5	\\
1992 A	&	1845	&	6	&	Jan 19.2$\pm$2(B)	&	Jan 1$\pm$3	&	...	&	...	&	...	\\
1992 G	&	1580	&	7	&	Feb 21.1$\pm$2(V)	&	Feb 1$\pm$5	&	...	&	...	&	...	\\
1993 L	&	1925	&	8	&	Apr 19$\pm$10	&	Apr 3$\pm$10	&	...	&	...	&	...	\\
1994 D	&	450	&	9	&	Mar 20.9$\pm$2 (B)	&	Mar 5$\pm$3	&	...	&	...	&	...	\\
1994 U	&	1329	&	10	&	Jul 5$\pm$7	&	Jun 15$\pm$8	&	{\it 940621}	&	{\it 2.21}	&	{\it 6$\pm$8}	\\
1994 ae	&	1282	&	11, 12	&	Nov 30$\pm$2	&	Nov 10$\pm$5	&	...	&	...	&	...	\\
1995 D	&	1967	&	 11-13	&	Feb 21.5$\pm$2	&	Feb 1$\pm$5	&	...	&	...	&	...	\\
1995 al	&	1541	&	14	&	Nov 7.1$\pm$2(B)	&	Oct 15$\pm$3	&	...	&	...	&	...	\\
1996 X	&	2032	&	14, 15	&	Apr 18$\pm$2(B)	&	Mar 31$\pm$3	&	...	&	...	&	...	\\
1996 Z	&	2275	&	14	&	May 13$\pm$2(B)	&	Apr 25$\pm$3	&	...	&	...	&	...	\\
1996 ai	&	992	&	14	&	Jun 20.8$\pm$2(B)	&	May 31$\pm$3	&	...	&	...	&	...	\\
1996 bk	&	2041	&	14	&	Oct 9.0$\pm$5(B)	&	Sep 25$\pm$7	&	{\it 960916}	&	{\it 2.81}	&	{\it 9$\pm$7}	\\
1996 bt	&	2675	&	16	&	Nov 1$\pm$7	&	Oct 11$\pm$8	&	{\it 961017}	&	{\it 2.89}	&	{\it 6$\pm$8}	\\
1997 bp	&	2492	&	10	&	Apr 7$\pm$7	&	Mar 18$\pm$8	&	...	&	...	&	...	\\
1997 bq	&	2813	&	17	&	Apr 17$\pm$2	&	Mar 30$\pm$3	&	970331	&	1.29	&	1$\pm$3	\\
1997 br	&	2069	&	18	&	Apr 20.3$\pm$2	&	Mar 30$\pm$3	&	...	&	...	&	...	\\
1997 dt	&	2191	&	19	&	Nov 22$\pm$7	&	Nov 2$\pm$8	&	...	&	...	&	...	\\
1998 aq	&	924	&	17	&	Apr 27$\pm$2	&	Apr 8$\pm$3	&	980406	&	1.04	&	2$\pm$3	\\
1998 bn	&	1828	&	20	&	Apr 30$\pm$5	&	Apr 10$\pm$7	&	...	&	...	&	...	\\
1998 bu	&	943	&	17	&	Apr 18$\pm$2	&	Mar 30$\pm$3	&	...	&	...	&	...	\\
1998 dg	&	2454	&	21, 22	&	Jul 30$\pm$10	&	Jul 10$\pm$10	&	...	&	...	&	...	\\
1998 dh	&	2669	&	17	&	Aug 2$\pm$2	&	Jul 15$\pm$3	&	{\it 980712}	&	{\it 2.88}	&	{\it 3$\pm$3}	\\
1998 dm	&	1943	&	23, 24	&	Aug 28$\pm$5	&	Aug 13$\pm$7	&	...	&	...	&	...	\\
1999 ac	&	2848	&	25	&	Mar 1$\pm$5	&	Feb 7$\pm$7	&	{\it 990206}	&	{\it 2.62}	&	{\it 1$\pm$7}	\\
1999 by	&	638	&	26, 27	&	May 11.25$\pm$2	&	Apr 27$\pm$5	&	...	&	...	&	...	\\
1999 cl	&	2120	&	28	&	Jun 15$\pm$2	&	May 26$\pm$5	&	...	&	...	&	...	\\
1999 cp	&	2823	&	28	&	Jun 18$\pm$5	&	May 29$\pm$7	&	...	&	...	&	...	\\
1999 gh	&	2310	&	29	&	Dec 3$\pm$7	&	Nov 13$\pm$8	&	...	&	...	&	...	\\
2000 E	&	1331	&	30	&	Feb 3$\pm$2	&	Jan 14$\pm$3	&	...	&	...	&	...	\\
2000 cm	&	2170	&	31, 32	&	May 22$\pm$10	&	May 2$\pm$10	&	{\it 000508}	&	{\it 2.52}	&	{\it 6$\pm$10}	\\
\enddata

\tablenotetext{a}{References. --- 1. Lira et al. 1998.  2. McNaught, Della Valle, \& Leisy 1991.  3. Leibundgut et al. 1993.  4. Filippenko et al. 1992.  5. Turatto et al. 1996.  6. Kirshner et al. 1993.  7. Ford et al. 1993.  8. Della Valle et al. 1993.  9. Vacca \& Leibundgut 1996.  10. Riess et al. 1998.  11. Ho et al. 2001.  12. Riess, Press, \& Kirshner 1996.  13. Sadakane et al. 1996.  14. Riess et al. 1999a.  15. Salvo et al. 2001.  16. Garnavich 1996.  17. Riess et al. 1999b.  18. Li et al. 1999.  19. Qiao et al. 1999.  20.  Patat \& Maia 1998.  21. Maza 1998.  22. Schmidt 1998.  23. Modjaz et al. 1998.  24. Filippenko \& De Breuck 1998.  25. Phillips, Kunkel, \& Filippenko 1999.  26. H\"oflich et al. 2002.  27. Toth \& Szabo 2000.  28. Krisciunas et al. 2000.  29. Nakano et al. 1999.  30. Vink\'o et al. 2001.  31. Jha, Challis, \& Kirshner 2000.  32. Turatto et al. 2000.}
    
\end{deluxetable}
\clearpage
\begin{figure}[!h]
\includegraphics[width=3.7cm,angle=270]{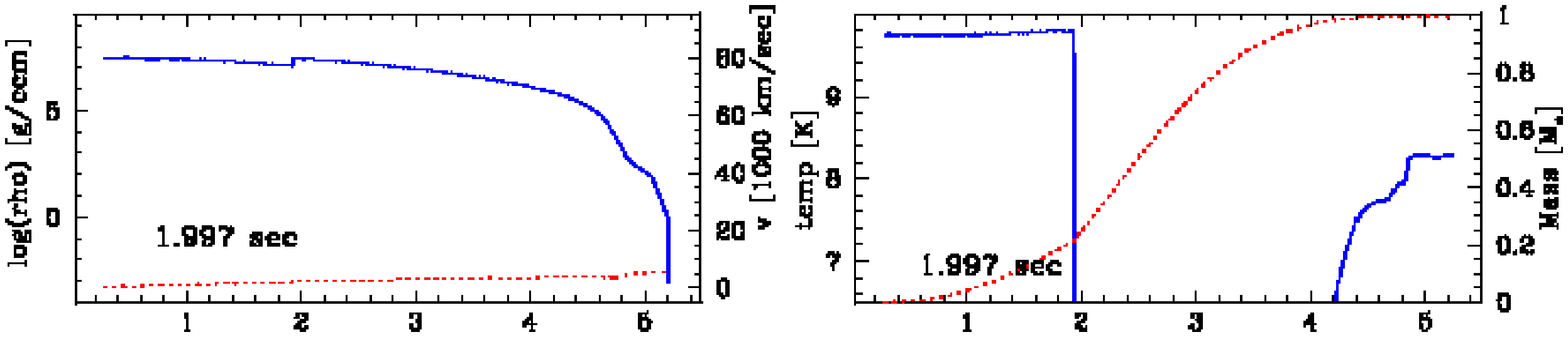}
\includegraphics[width=3.7cm,angle=270]{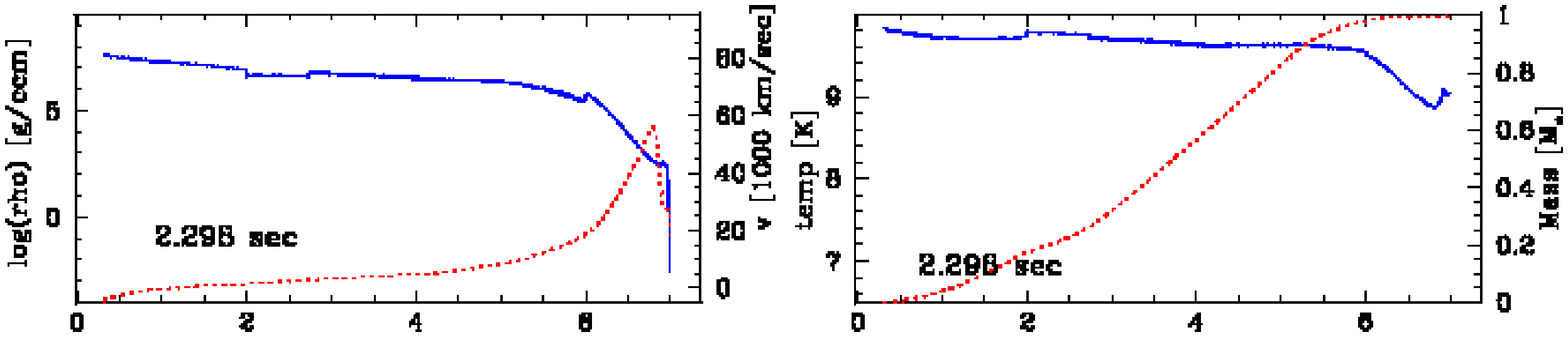}
\includegraphics[width=3.7cm,angle=270]{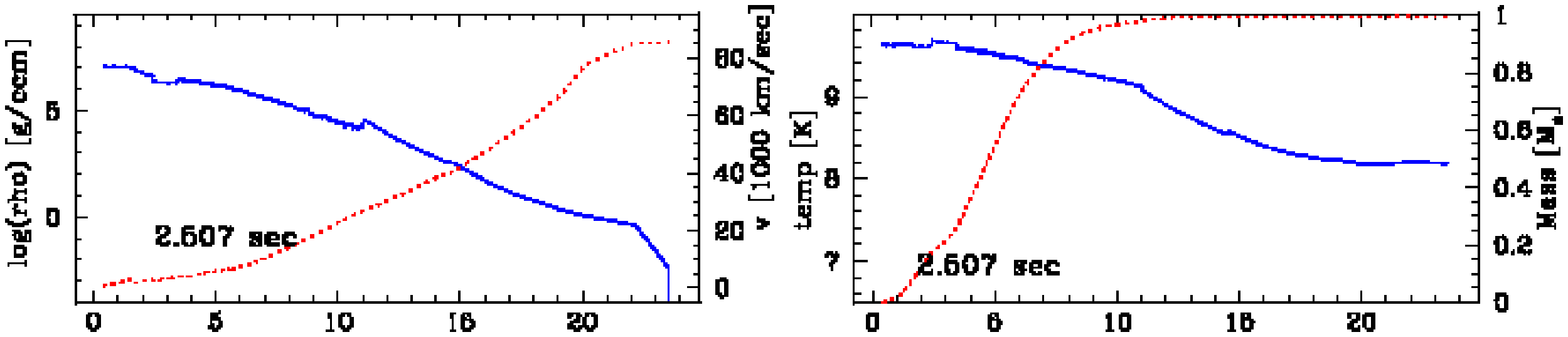}
\includegraphics[width=3.7cm,angle=270]{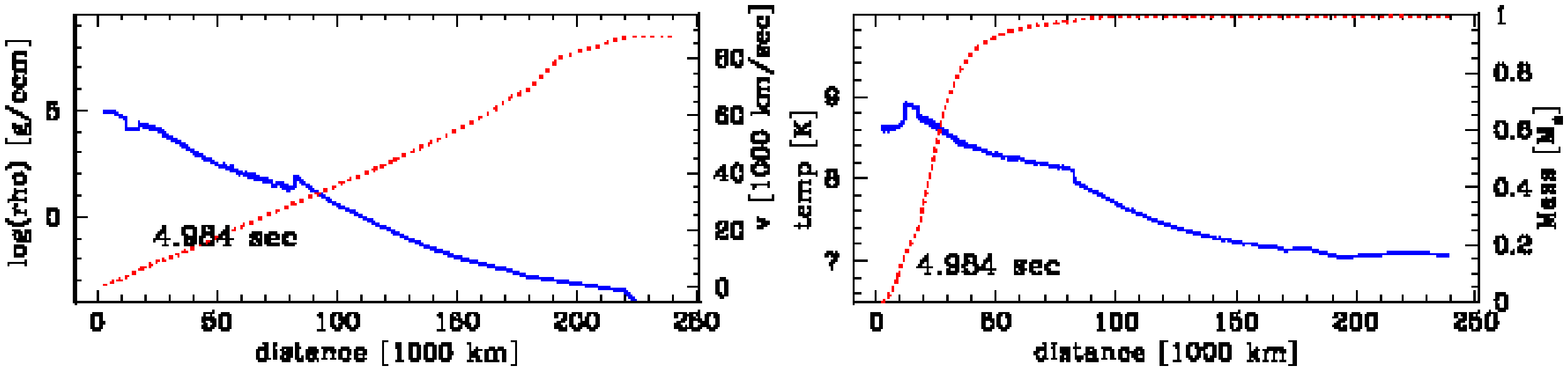}
\caption{
Structure of the delayed detonation model as a function of distance at various 
stages of the explosion, namely a) during the deflagration phase, b) at the 
shock breakout, c) during the strong acceleration phase of
the outer layers, and d) close to homologous expansion.  We give the density 
(solid, left scale), velocity
(dotted, right scale), and the logarithm of the temperature (solid, left scale),
 enclosed mass (dotted, right scale), respectively.
}
\label{time}
\end{figure}

\clearpage

\begin{figure}[!h]
\includegraphics[width=12.7cm,angle=270]{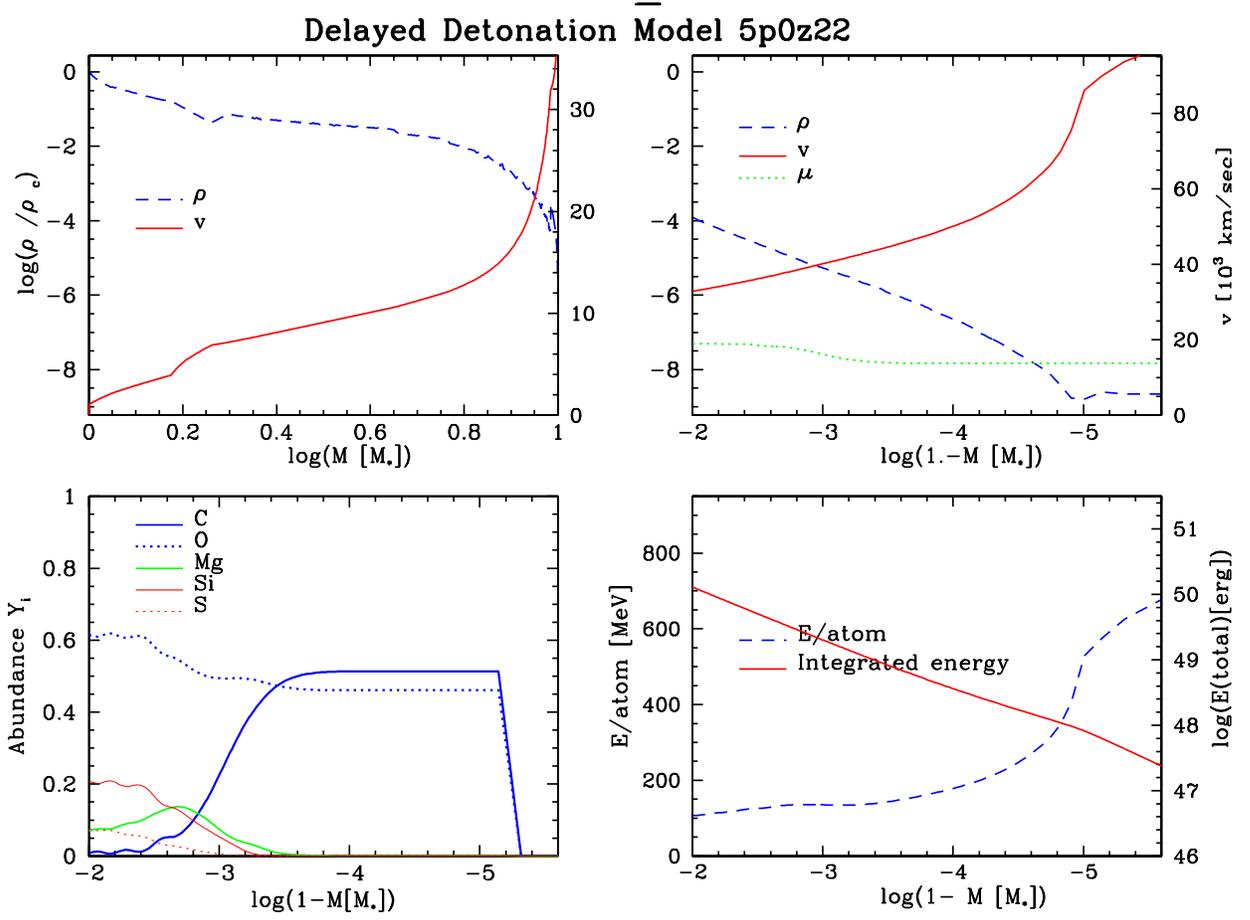}
\caption{The density, velocity,  and mean molecular weight at about 10 seconds
after the explosion (upper plots) are given with the scale on the left and right, 
respectively. At this time, the expansion is close to homologous.  On the lower 
left, the  chemical abundances for the outer $10^{-2}M_\odot $. For the full 
profiles, see H\"oflich et al. (2002).  Outside $5 \times 10^{-6} M_\odot $ the abundance
is solar. In addition, the mean energy per atom (right scale) and the integrated 
kinetic energy (from the surface) are given on the lower right.}
\label{rho}
\end{figure}

\clearpage

\begin{figure}[!h]
\includegraphics[width=12.7cm,angle=270]{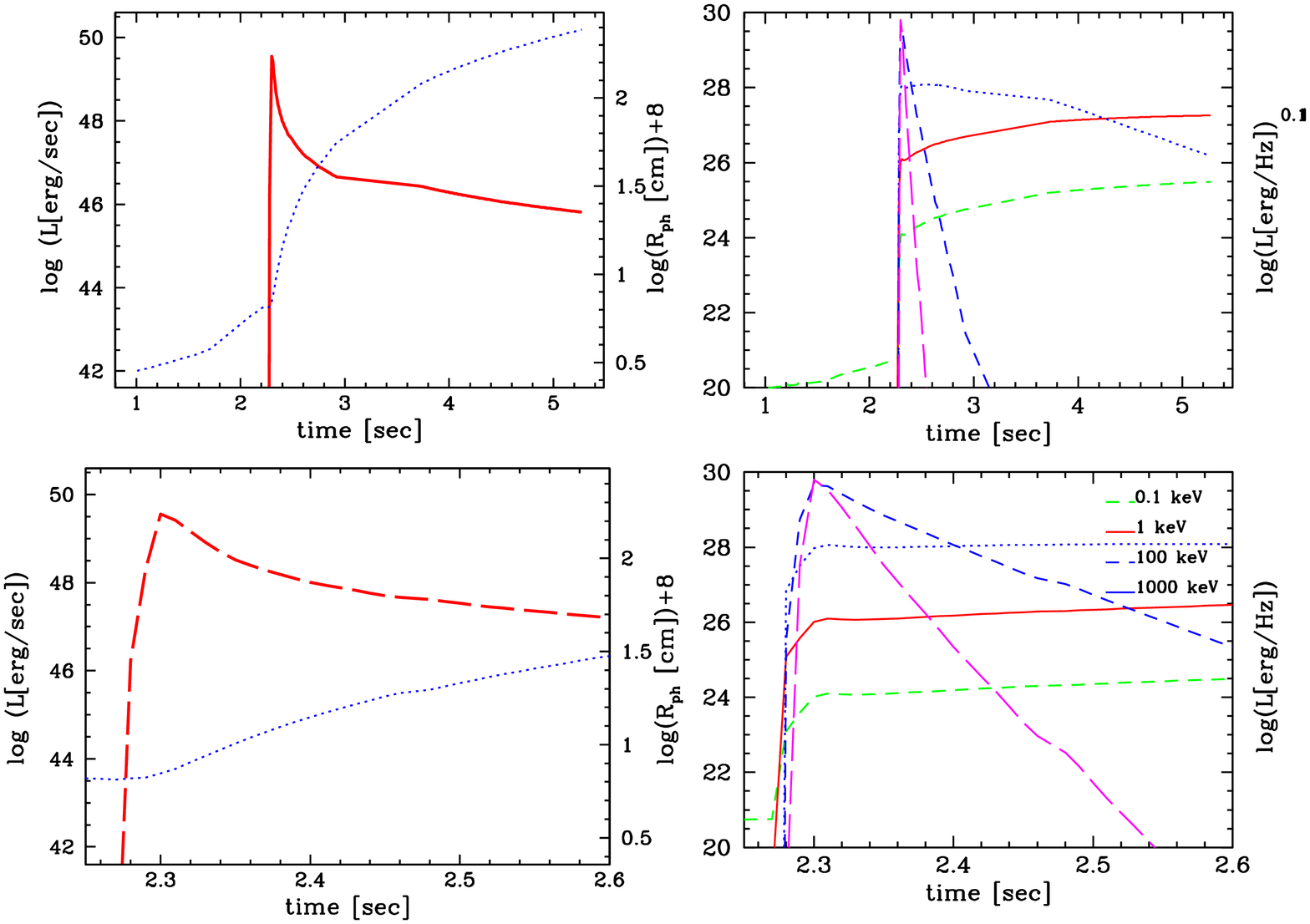}
\caption{Bolometric and monochromatic luminosities as a function of time. In addition, the
evolution of the photospheric radius is given which is within the outer $10^{-4}M_\odot $
during at the first minute.
}
\label{lum}
\end{figure}


\begin{thebibliography}{}

\bibitem[]{} Barbon, R., Buond, V., Cappellaro, E., \& Turatto, M. 1999, A\& AS, 139, 531; see also http://web.pd.astro.it/supern/snean.txt.
\bibitem[]{} BATSE GRB Team, 2001, http://gammaray.msfc.nasa.gov/batse/grb/catalog/current/.
\bibitem[Benz et al.(1990)]{benz90} Benz, W. Cameron, A. G. W., Press, W. H. \& Bowers, R. L. 1990,        ApJ, 348, 647
\bibitem[]{} Berger, E. et al. 2005, Nature, 438, 988.
\bibitem[]{} Bloom, J. S. et al. 1999, Nature, 401, 453.
\bibitem[]{} Bloom, J. S. et al. 2006, ApJ, 638, 354.
\bibitem[Branch(1999)]{branch99} Branch, D. 1999, ARAA, 36, 17
\bibitem[]{} Briggs, M. S. et al. 1999, ApJSupp, 122, 503.
\bibitem[]{} Briggs, M. S. 2001, in Gamma- Ray Bursts in the Afterglow Era, ed. E. Costa, F. Frontera, and J.  Horth (Berlin: Springer), 22.
\bibitem[]{} Burrows, D. N. et al. 2005, GCN Circ. 4366.
\bibitem[]{} Campana, S. et al. 2006, Nature, 442, 1008.
\bibitem[]{} Cline, T. L. \& Desai, U. 1974, in Proc. 9th ESLAB Symp. (Noordwijk: ESRO), 37.
\bibitem[]{} Cline, T. L. et al. 1980, ApJ, 237, L1.
\bibitem[]{} Colgate, S. A. 1968, Canadian J. Phys., 46, S476.
\bibitem[]{} Colgate, S. A. 1970, Acta Physica Academiae Scientiarum Hungaricae, 29, Suppl. 1, pp. 353-359.
\bibitem[]{} Colgate, S. A. 1974, ApJ, 187, 333.
\bibitem[]{} Dado, S. \& Dar, A. 2005, GCN Circ. 3423.
\bibitem[]{} Dar, A. 1999, A\&AS, 138, 505.
\bibitem[]{} Dar, A. \& De Rujula, A. 2004, Physics Reports, 405, 203.
\bibitem[]{} Della Valle, M. et al. 1993, IAUCirc. 5782.
\bibitem[]{} Deng, M. 2001, Ph. D. Thesis, Yale University.
\bibitem[]{} Evans, W. D. et al. 1980, ApJ, 237, L7.
\bibitem[]{} Felten, J. E. 1982, 17th International Cosmic Ray Conference, Paris, 9, 52-55.
\bibitem[]{} Fenimore, E. E. \& Ramirez-Ruiz, E. 2000, astro-ph/0004176.
\bibitem[]{} Filippenko, A. V. \& De Breuck, C. 1998, IAUCirc. 6997.
\bibitem[]{} Filippenko, A. V. et al. 1992, AJ, 104, 1543.
\bibitem[Fisher et al.(1999)]{fisher99} Fisher  A., Branch  D., Hatano  K., 	 Baron  E. 1999,   MNRAS, 304, 679
\bibitem[]{} Fishman, G. J. et al. 1994, ApJSupp, 92, 229.
\bibitem[]{} Ford, C. H. et al. 1993, AJ, 106, 1101.
\bibitem[]{} Fox, D. B. et al. 2005, Nature, 437, 845.
\bibitem[]{} Galama, T. J. et al. 1998, Nature, 395, 670.
\bibitem[]{} Galama, T. J. et al. 2000, ApJ, 536, 185.
\bibitem[]{} Garnavich, P. 1996, IAUCirc. 6503.
\bibitem[]{} Gehrels, N. et al. 2004, ApJ, 611, 1005.
\bibitem[]{} Gehrels, N. et al. 2005, Nature, 437, 851.
\bibitem[Gerardy et al(2004)]{gerardy04} Gerardy, C. L., H\"oflich, P., Fesen, R. A., Marion, G. H., Nomoto, K., Quimby, R., Schaefer, B. E., Wang, L., \& Wheeler, J. C. 2004, ApJ, 607, 1258
\bibitem[Hatano et al.(1999)]{Hatano99} Hatano, K., Branch, D., Fisher, A., Baron, E., \& Filippenko, A. V. 1999, ApJ, 525, 881
\bibitem{hatano00} Hatano  K., Branch  D., Lentz  E. J., 	Baron  E.,  Filippenko  A. V.,  Garnavich  P. 2000, ApJ 543L, 49
\bibitem[]{} Heise, J., in't Zand, J., Kippen, R. M., \& Woods, P. M. 2001, in Gamma-Ray Bursts in the Afterglow Era, ed. E. Costa, F. Frontera, and J. Horth (Berlin: Springer), 16.
\bibitem[]{} Hjorth, J. et al. 2003, Nature, 423, 847.
\bibitem[]{} Hjorth, J. et al. 2005, Nature, 437, 859.
\bibitem[]{} Ho, W. C. G. et al. 2001, PASP, 113, 1349.
\bibitem[H\"oflich(2006)]{hoeflich05} H\"oflich P. 2005, Nuclear Physics A, 777, 579
\bibitem[H\"oflich et al.(2005)]{h05} H\"oflich P., Gerardy C., Quimby R., 2005, New Astronomy, 50, 470
\bibitem[H\"oflich et al. (2004))]{hetal04} H\"oflich, P., Gerardy C., Nomoto K., Motohara K., Fesen R.A., Maeda K., Ohkubo T., Tominaga N, ApJ, 617, 1258
\bibitem[]{hetal03} H\"oflich, P., Gerardy, C., Linder, E., \& Marion, H.  	2003,   Lecture Notes in Physics 635, Springer Press, p. 203 \& astro-ph/0301334
\bibitem[H\"oflich et al.(2002)]{h02} H\"oflich, P., Gerardy, C., Fesen, R., \& Sakai, S. 2002,   	ApJ, 568, 791
\bibitem[H\"oflich(2002a)]{hoeflich02a} H{\" o}flich P., 2002b, in: Theory of  Stellar Atmospheres, eds. I. Hubeney, D. Mihalas
 \& C. Werner, ASP Conference Series,  Vol. 288, p. 371
\bibitem[H\"oflich(2002b)]{hoeflich02b} H{\" o}flich P., 2002b, in: Theory of  Stellar Atmospheres, eds. I. Hubeney, D. Mihalas
 \& C. Werner, ASP Conference Series,  Vol. 288, p. 185
\bibitem[]{} H\"oflich, P., Gerardy, C., Fesen, R. A., \& Sakai, S. 2002, ApJ, 568, 791.
\bibitem[H\"oflich \& Khokhlov (1996)]{hk96} H\"oflich  P.,  Khokhlov  A. 1996,         ApJ, 457, 500
\bibitem[hetal98]{hwt98} H\"oflich  P.,         Wheeler  J. C.,  Thielemann  F. K. 1998,  ApJ, 495, 617
\bibitem[H\"oflich(1995)]{h95} H\"oflich  P. 1995,  ApJ, 443, 89
\bibitem[H\"oflich et al. (1991)]{hoeflich93} H\"oflich  P., 	Khokhlov  A.,  M\"uller E.   1993, A\&A, 270, 223
\bibitem[]{} Hurley, K. 2000, in Gamma-Ray Bursts: 5th Huntsville Symposium, ed. R. M.  Kippen, 	R. S. 	Mallozzi, and G. J. Fishman (Melville NY, AIP Conf. Proc. 526), 763. 
\bibitem[]{} Hurley, K. et al. 1999, ApJSupp, 120, 399.
\bibitem[]{} Hurley, K. et al. 2000, ApJ, 534, 258; see also http://ssl.berkeley.edu/ipn3/index.html.
\bibitem[]{} Hurley, K. et al. 2005, Nature, 434, 1098.
%\bibitem[Iben(1975)]{iben75} Iben Jr  I., Tutukov  A.V. 1975, 	ApJs, ??, ??
\bibitem[]{} Jha, S., Challis, P., \& Kirshner, R. 2000, IAUCirc. 7437.
\bibitem[]{} Kahabka, P., van den Heuvel, E.P.J. 1997, A\&A 35, 69
\bibitem[]{} Kippen, R. M. et al. 1998, ApJ, 506, L27.
\bibitem[]{} Kippen, R. M., Woods, P. M., Heise, J., in't Zand, J., Preece, R. D., \& 
\bibitem[Khokhlov(1991)]{khokhlov91} Khokhlov  A. 1991, ApJ, 245, 114
\bibitem[]{} Kirshner, R. P. 1993, ApJ, 415, 589.
\bibitem[]{} Klebesadel, R. W., Strong, I. B., \& Olson, R. A. 1973, ApJ, 182, L85.
\bibitem[]{} Kouveliotou, C. et al. 1993, ApJ, 413, L101.
\bibitem[]{} Kouveliotou, C. et al. 1993, ApJ, 567, 447.
\bibitem[]{} Krisciunas, K. et al. 2000, ApJ, 539, 658.
\bibitem[]{} Kulkarni, S. et al. 1998, Nature, 395, 663.
\bibitem[]{} Leibundgut, B. et al. 1993, AJ, 105, 301.
\bibitem[]{} Li, W. D. et al. 1999, AJ, 117, 2709.
\bibitem[]{} Li, W. D. et al. 2003, PASP, 115, 453.
\bibitem[]{} Lira, P. et al. 1998, AJ, 115, 234.
\bibitem[]{} MacFadyen, A. I., Woosley, S. E., \& Heger, A. 2001, ApJ, 550, 410.
\bibitem[]{} Malesani, D. et al. 2004, ApJLett, 609, L5.
\bibitem[Marion, H{\" o}flich, Vacca, \& Wheeler(2003)]{marion03}Marion, G.~H., H{\" o}flich, P., Vacca, W.~D., \& Wheeler, J.~C.\ 2003, ApJ, 591, 316
\bibitem[]{} Maza, J. 1998, IAUCirc. 6978.
\bibitem[]{} Marietta E., Burrows A., Fryxell B. 2000, ApJ.Suppl. 128, 615
\bibitem[Mattila et al. (2005)]{mattila05} Mattila et al. 2005, A\&A, 443, 649
\bibitem[]{} McNaught, R. N., Della Valle, M., \& Leisy, P. 1991, IAUCirc. 5258.
\bibitem[]{} Modjaz, M. et al. 1998, IAUCirc. 6993.
\bibitem[]{} Nakano, S. et al. 1999, IAUCirc. 7328.
\bibitem[Nomoto et al.(2003)]{nomoto03} Nomoto, K., Uenishi, T., Kobayashi, C., Umeda, H., Ohkubo,  T., Hachisu, I., \& Kato, M. 2003, in: in From Twilight to Highlight: The Physics of Supernovae, eds. W. Hillebrandt \& B. Leibundgut, ESO Astrophysics Symposia (Berlin:Springer), 115  (astro-ph/0308138)
\bibitem[]{} Norris, J. P., Marani, G., \& Bonnell, J. 2000, ApJ, 534, 248.
\bibitem[]{} Paciesas, W. C. et al. 1999, ApJSupp, 122, 465.
\bibitem[]{} Pain, R. et al. 1996, ApJ, 473, 356.
\bibitem[]{} Patat, F. \& Maia, M. 1998, IAUCirc. 6888.
\bibitem[]{} Phillips, M. M., Kunkel, W., \& Filippenko, A. V. 1999, IAUCirc. 7122.
\bibitem[]{} Pozdnyakov L.A., Sobol I.M., Syunyaev R.A., 1976, Soviet Astr. Let. 2, 55
\bibitem[Quimby et al.(2005)]{q05} Quimby R., H\"oflich P., Kannapa S.J., Rykoff E., Rujopakarn W., Akerlof C.W.,  Gerardy C., Wheeler J.C. 2005, ApJ, in press
\bibitem[]{} Qiao, Q. Y., Qiu, Y. L., Li, W. D., \& Hu, J. Y. 1999, IAUCirc. 6775.
\bibitem[]{} Rappaport S/. Chiang, E., Kallman, T., Malina, R. 1994, ApJ 431, 237
\bibitem[]{} Reichart, D. E. 1999, ApJ, 521, L111.
\bibitem[]{} Riess, A. G., Nugent, P., Filippenko, A. V., Kirshner, R. P., \& 
\bibitem[Paczy\'nski(1985)]{pac85} Paczy\'nski  B., 1985, in: Cataclysmic        Variables and Low-Mass X-Ray Binaries, eds. D.Q. Lamb, J. Patterson, Reidel, Dordrecht, p.1
\bibitem[]{} Perlmutter, S. 1998, ApJ, 504, 935.
\bibitem[]{} Riess, A. G., Press, W. H., \& Kirshner, R. P. 1996, ApJ, 473, 88.
\bibitem[]{} Riess, A. G. et al. 1999a, AJ, 117, 707.
\bibitem[]{} Riess, A. G. et al. 1999b, ApJ, 118, 2675.
\bibitem[]{} Sadakane, K. et al. 1996, PASJ, 48, 51.
\bibitem[]{} Salvo, M. E. et al. 2001, MNRAS, 321, 254.
\bibitem[]{} Schaefer, B. E. 1996, ApJ, 464, 404.
\bibitem[]{} Schaefer, B. E. 2006, ApJLett, 642, L25.
\bibitem[]{} Schaefer, B. E. \& Deng, M. 2000, in Gamma-Ray Bursts: 5th Huntsville Symposium, ed. R. M. Kippen, 	R. S. Mallozzi, and G. J. Fishman (Melville NY, AIP Conf. Proc. 526), pp. 419-423.  Schaefer, B. E. 2003, ApJ, 583, L71.
\bibitem[]{} Schaefer, B. E. et al. 2003, ApJ, 588, 387.
\bibitem[]{} Schmidt, B. 1998, IAUCirc. 6989.
\bibitem[]{} Soderberg, A. M. et al. 2006, ApJ, 650, 261.
\bibitem[]{} Stanek, K. Z. et al. 2003, ApJ, 591, L17.
\bibitem[]{} Taylor, J.H 1994, Rev. Mod. Phys. 66, 711
\bibitem[]{} Terlevich, R. \& Fabian A. 1999, IAUCirc. 7269.
\bibitem[]{} Toth, I. \& Szabo, R. 2000, A\&A, 361, 63.
\bibitem[]{} Turatto, M., Pastorello, A., Cappellaro, E., \& Cedrati, F. 2000, IAUCirc.  7438.
\bibitem[]{} Turatto, M. et al. 1996, MNRAS, 283, 1.
\bibitem[]{} Vacca, W. D. \& Leibundgut, B. 1996, ApJ, 471, L37.
\bibitem[]{} Van den Heuvel, E.P.J., Bhattacharya, D., Nomoto, K., Rappaport, S. 1992, A\&A 262, 97
\bibitem[]{} Villasenor, J. S. et al. 2005, Nature, 437, 855.
\bibitem[]{} Vink, J. et al. 2001, A\&A, 372, 824.
\bibitem[]{} van den Bergh, S. \& Tammann, G. A. 1991, ARA\&A, 29, 363.
\bibitem[]{} Waxman E.; Meszaros P., Campana S. 2007, ApJ, accepted, \& arXiv:astro-ph/0702450
\bibitem[]{} Wang, L. \& Wheeler, J. C. 1998, ApJ, 504, L87.
\bibitem[Wang et al.(2003)]{wang03}  Wang, L., Baade, D., H\"oflich, P., Khokhlov, A, Wheeler, J. C., Kasen,  D., Nugent  P., Perlmutter S., Fransson C., Lundqvist P. 2003, ApJ 591, 1110
\bibitem[Wang et al.(2006)]{wang05}  Wang, L., Baade, D., H\"oflich, P., Wheeler J.C., Kawabata, K., Khokhlov, A,
Nomoto, K., Patat, F.2006 ApJ, 653, 490
\bibitem[Webbing(1984)]{webbink94} Webbink  R. F. 1984, ApJ, 277, 355
\bibitem[Whelan(1973)]{Whelan73} Whelan  J.,  Iben Jr.  I. 1973, ApJ,    	186, 1007
\bibitem[]{} Woosley, S. E. \& Bloom, J. S. 2006, ARA\&A, 44, 507.

\end{thebibliography}
\end{document}